# Supernova-Remnant Origin of Cosmic Rays?

Yousaf M. Butt*, Diego F. Torres#, Gustavo E. Romero†, Thomas Dame*, Jorge A. Combi†

*Harvard-Smithsonian Center for Astrophysics, 60 Garden St., Cambridge, MA, USA.

#Department of Physics, Joseph Henry Laboratories, Jadwin Hall, P.O. Box 708, Princeton University, Princeton,NJ 08544-0708, USA.

†Instituto Argentino de Radioastronomía, CC 5, 1894, Villa Elisa, Buenos Aires, ARGENTINA.

**It is thought that Galactic cosmic ray (CR) nuclei are gradually accelerated to high energies (up to ~300 TeV/nucleon, where 1TeV=$10^{12}$eV) in the expanding shock-waves connected with the remnants of powerful supernova explosions. However, this conjecture has eluded direct observational confirmation[1,2] since it was first proposed in 1953 (ref. 3). Enomoto et al.[4] claim to have finally found definitive evidence that corroborates this model, proposing that the very-high-energy, TeV-range, $\gamma$-rays from the supernova remnant (SNR) RX J1713.7-3946 are due to the interactions of energetic nuclei in this region. Here we argue that their claim is not supported by the existing multiwavelength spectrum of this source. The search for the origin(s) of Galactic cosmic ray nuclei may be closing in on the long-suspected supernova-remnant sources, but it is not yet over.**

We have previously suggested[5] that the SNR RX J1713.7-3946 (dark X-ray contours in Fig. 1), might be an accelerator of CR nuclei on the basis of GeV ($10^9$ eV) $\gamma$-ray emission[6] (Fig. 1, white contours) seen towards the adjacent massive and dense molecular clouds (Fig. 1, false colour) with which it seems to be interacting[7]; the $\gamma$-ray signature of nucleonic interactions is known to be greatly amplified in such dense media. Enomoto et al. have since claimed[4] that the $\gamma$-rays of even higher energy (TeV range; red contours in Fig. 1) that they detected directly towards the more rarefied remnant are an unambiguous signature of nucleonic interactions in this region, thus finally proving the long-held conjecture.

However, it is well-known that SNRs accelerate electrons (see ref. 8, for example) that can also generate TeV $\gamma$-rays at the source sites. It is therefore generally difficult to assign conclusively a nucleonic rather than a electronic origin to the observed celestial $\gamma$-rays[9]. Fortunately, in this case the discrimination is not so subtle: if the TeV emission seen in the northwest quadrant were really due to protons, as proposed[4], that part of the SNR would have been detected as a bright GeV source in the Energetic Gamma Ray Experiment Telescope (EGRET) all-sky survey[6], but it was not.

The proposed model[4] is unsatisfactory because it predicts that there will be roughly three times the GeV gamma-ray intensity of a nearby EGRET source[6] (3EG 1714-3857) at a location where no actual GeV flux has been detected (that is, above the background level), although alternative models may reduce this discrepancy (see also ref. 10). In addition, the high gas density (about $10^2$ protons $cm^{-3}$) needed to explain the TeV gamma-rays as nucleonic in origin[4] is simply not compatible – by several orders of magnitude – with the X-ray data[7] from the same region of the SNR

As the peaks of both the TeV (ref. 4) and X-ray (ref. 7) emissions lie in the same low-density region[7] of the remnant (Fig. 1]), the most plausible explanation is that they both arise from the same population of relativistic electrons. However, there remains an exciting possibility that a minor component of the reported TeV emission[4] could be due to nucleonic interactions at the location of the adjacent Cloud B (Fig. 1), where sufficient molecular material is present[5]. A suggestive extension and a local maximum in the TeV significance contours[4] are coincident with this location (Fig. 1).

SNR RX J1713.7-3946 might therefore be accelerating nuclei to CR energies[5], but the TeV gamma-ray signature so far detected from the northwest quadrant of the remnant is not proof of this. It is likely that most – but not all – of the observed TeV emission is actually electronic in origin.

To address the question of the source(s) of the TeV emission properly, better radio data sampled at several frequencies, as well as the inclusion of existing low-energy ROSAT X-ray data, will be needed to determine the primary synchrotron spectrum. Data of greater spatial resolution from the next-generation GeV, TeV and neutrino telescopes will also be important. Only when this multifrequency data set is available can the nature of the contributions (nucleonic *vs.* electronic) to the observed TeV emissions begin to be disentangled.

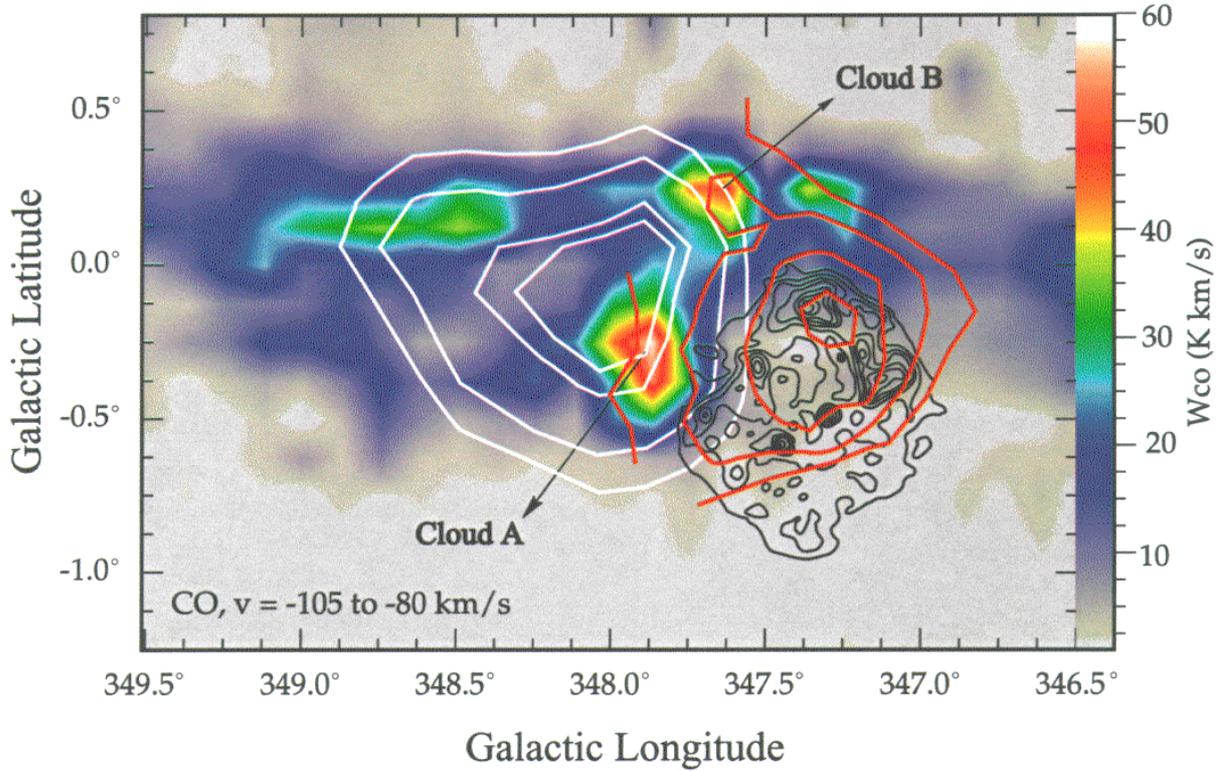

Fig 1: An overlay map in Galactic coordinates showing SNR RX J1713.7-3946 (G347.3-0.5) in dark contours (*ROSAT* PSPC X-ray) from Ref 7. Red depicts the TeV significance contours (Ref 4). In white are the location probability contours of the GeV *EGRET* source 3EG J1714-3857 (Ref 6). The color-scale indicates the intensity of CO($J$=1→0) emission, and consequently the column density of the ambient molecular cloud, in the LSR velocity interval $v_{lsr}$= -105 to -80 km/sec associated with the SNR, corresponding to a kinematic distance of 6.3±0.4 kpc (Ref. 5). Note the local maximum in the TeV significance contours at Cloud B, which may indicate a hadronic origin of the TeV flux *at that location*. For further details please see Ref 5.

**Correspondence and requests for materials should be addressed to Y.M.B. (e-mail: ybutt@cfa.harvard.edu)**

*We are very grateful to Olaf Reimer & Martin Pohl – who independently and simultaneously arrived at essentially the same conclusions described here (see Ref. 10) – for very useful and open discussions. We also thank Glenn Allen, Thomas Panutti, and Pat Slane for valuable comments.*